\documentclass[11pt,twoside]{article}
\usepackage[pdftex]{graphicx}
\usepackage{amsmath}
\usepackage{amssymb}
\usepackage{cite}

\begin{document}
\newcommand{\pst}{\hspace*{1.5em}}

\newcommand{\rigmark}{\em Journal of Russian Laser Research}
\newcommand{\lemark}{\em Volume 30, Number 5, 2009}

\newcommand{\be}{\begin{equation}}
\newcommand{\ee}{\end{equation}}
\newcommand{\bm}{\boldmath}
\newcommand{\ds}{\displaystyle}
\newcommand{\bea}{\begin{eqnarray}}
\newcommand{\eea}{\end{eqnarray}}
\newcommand{\ba}{\begin{array}}
\newcommand{\ea}{\end{array}}
\newcommand{\arcsinh}{\mathop{\rm arcsinh}\nolimits}
\newcommand{\arctanh}{\mathop{\rm arctanh}\nolimits}
\newcommand{\bc}{\begin{center}}
\newcommand{\ec}{\end{center}}

\thispagestyle{plain}

\label{sh}


\begin{center} {\Large \bf
\begin{tabular}{c}
NONLINEAR CHANNELS OF WERNER STATES.
\end{tabular}
 } \end{center}

\bigskip

\bigskip

\begin{center} {\bf
V.I. Man'ko$^{1*}$ and R.S. Puzko$^2$
}\end{center}

\medskip

\begin{center}
{\it
$^1$P.N. Lebedev Physical Institute, Russian Academy of Science\\
Moscow, Russia 119991

\smallskip

$^2$Moscow Institute of Physics and Technology (State University)\\
Dolgoprudny, Russia 117303
}
\smallskip

$^*$Corresponding author e-mail:~~~manko~@~sci.lebedev.ru\\
\end{center}

\begin{abstract}\noindent
The nonlinear positive map of density matrix of two-qubit Werner state called nonlinear channel is studied. The map $\rho\to\Phi(\rho)$ is realized by “rational” function $\Phi$. The influence of the map onto the entanglement properties of the transformed density matrix is discussed. The violation of Bell inequality (CHSH inequality) for the two-qubit state $\Phi(\rho)$ is investigated. The nonlinear channels under discussion create the entangled state from separable Werner state. The quantum spin-tomograms of the states are studied.
\end{abstract}

\medskip


\section{Introduction}
\pst
The entanglement~\cite{1} of composite quantum system states is related to purely quantum correlations of the subsystems, e.g. for two-qubit states this phenomenon is studied for Werner states~\cite{2}. The Peres-Horodecki criterion~\cite{3,4,5} of the entanglement is known as necessary and sufficient condition for two-qubit state and for qubit-qutrit state. The density matrix of any system state can be transformed into another density matrix. Linear transforms of density matrix are known as positive maps~\cite{6} (see also~\cite{7}). Some specific positive maps are known as completely positive maps~\cite{6,8} called quantum channels. Recently some nonlinear maps of density matrices were discussed~\cite{9,10}. In~\cite{15,16} the spin-tomograms of qudit states were introduced. The tomograms are fair probability distributions which determine the density matrix of the spin-system state.

The aim of our work is to study how the nonlinear positive map (nonlinear channel) influences the entanglement properties, if it is applied to the Werner state. We will discuss the properties of the spin-tomogram of Werner state under the action of nonlinear channels. The standard linear quantum channels are intensively discussed in the literature (see e.g.\cite{12}).We consider the specific nonlinear map. This map provides the new density matrix which is obtained from the initial one by taking $n$-th power of this initial matrix with adding the corresponding factor responsible for normalization condition of the density operator. The important property of the entangled qudit-system state is the possibility to violate Bell inequality (CHSH). We are going to study in this work the influence of nonlinear channels on the violation of CHSH inequality.

The paper is organized as follows.

In Sec.~2 we review the properties of Werner state and its density matrix $\rho_{w,1}$ which depends on the real parameter p. The entanglement phenomenon is studied for density matrix $\rho_{w,1}^n/Tr[\rho_{w,1}^n]$ is Sec.~3. The tomograms and violation of Bell inequalities for two series of transformed (by nonlinear channels) states are discussed in Sec.~4. The conclusion and prospective are presented in Sec.~5.

\section{Properties of non-linear channels $\rho_{w,1}^n/Tr[\rho_{w,1}^n]$ for cases $n=1,2,3$}
\pst
Density matrix for Werner states with parameter $p$ can be written as
\be
\rho_{w,1}=\begin{pmatrix}
A & 0 & 0 & C\\
0 & B & 0 & 0\\
0 & 0 & B & 0\\
C & 0 & 0 & A\\
\end{pmatrix},
\ee
where $A(p)=\frac{1+p}{4}$, $B(p)=\frac{1-p}{4}$, $C(p)=\frac{p}{2}$. Eigenvalues for this matrix are 
\be
\lambda_1=\frac{1+3p}{4},\lambda_{2,3,4}=\frac{1-p}{4}. 
\ee
\begin{figure}[ht]
\bc \includegraphics[width=8.6cm]{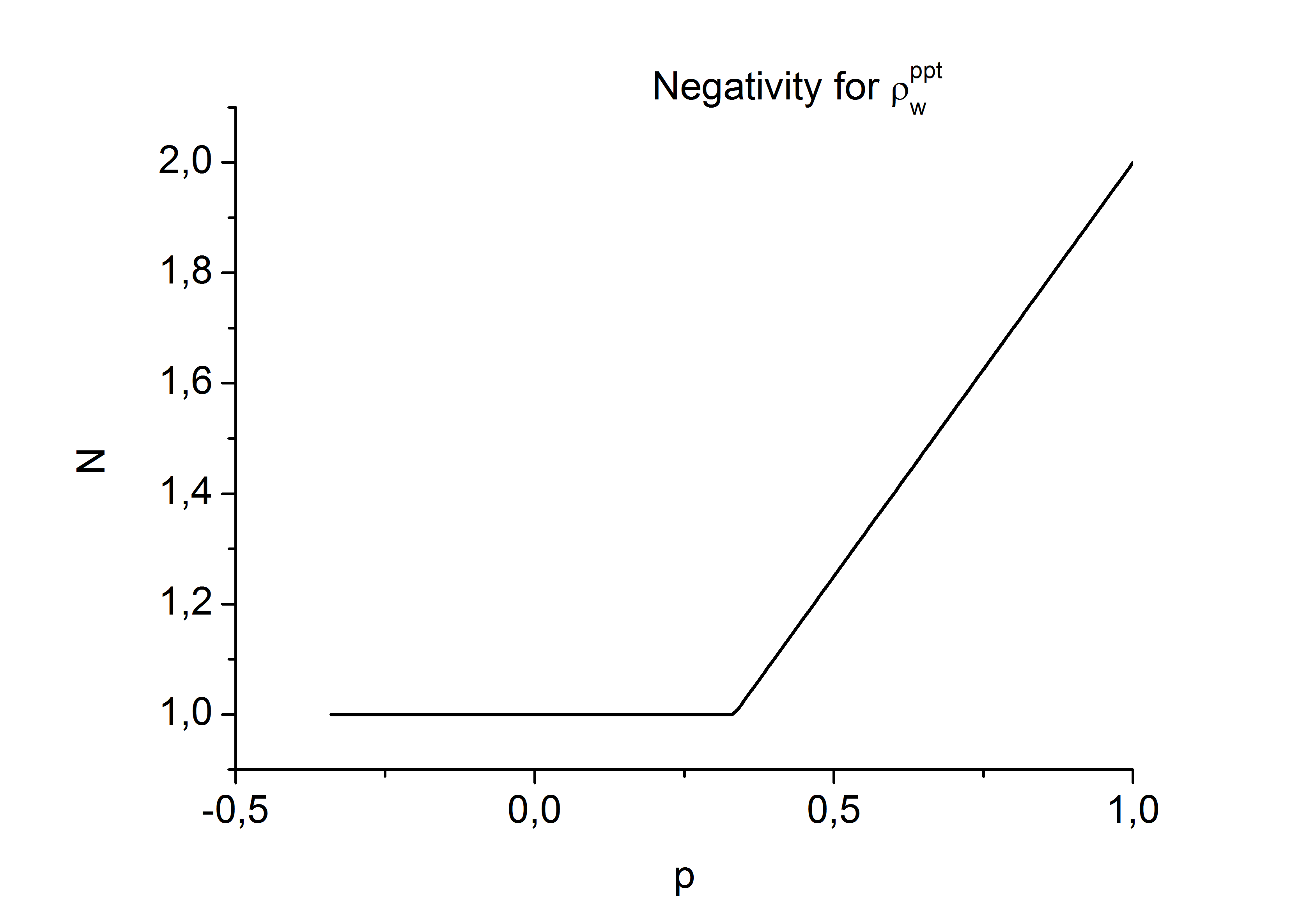}
\includegraphics[width=8.6cm]{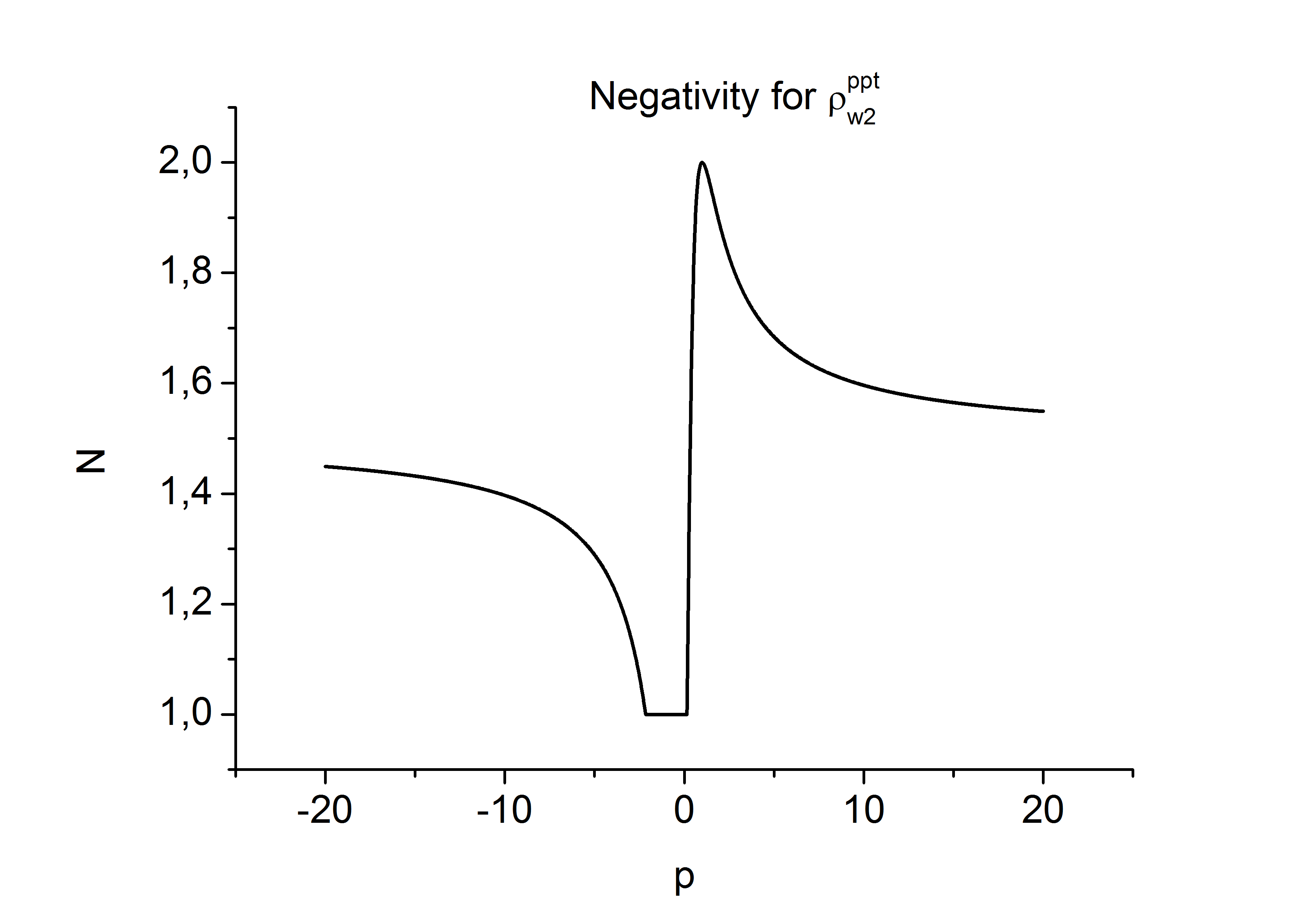} \ec
\bc 
\includegraphics[width=8.6cm]{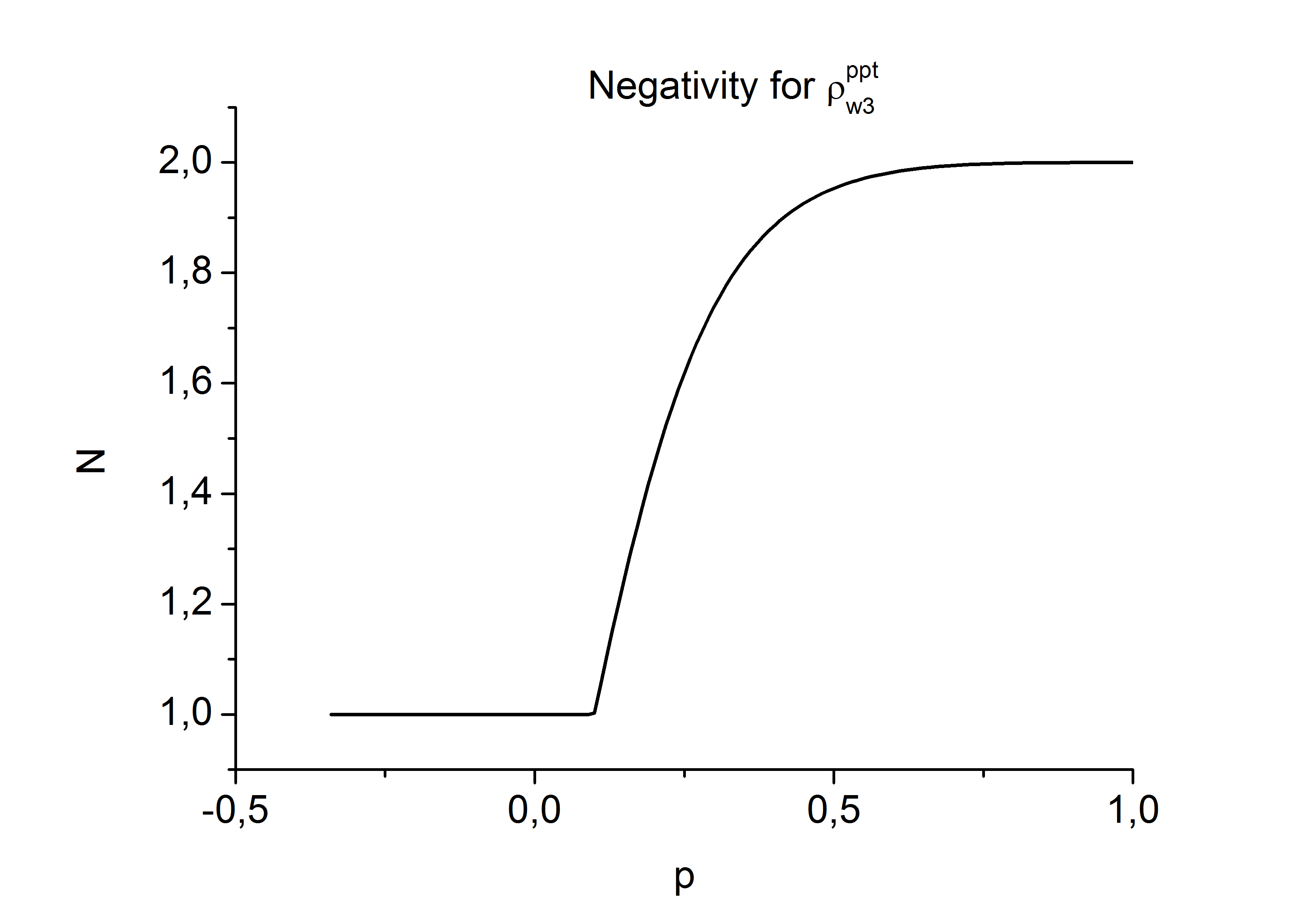} 
\ec
\vspace{-4mm}
\caption{Negativity (the sum of absolute values of eigenvalues of density matrix) for cases $\rho_{w,1}^{ppt}$ (upper-left plot), $\rho_{w,2}^{ppt}$ (upper-right plot) and $\rho_{w,3}^{ppt}$ (lower plot).}
\end{figure}
Non-negativity of density matrix restricts parameter $p$ to domain $-\frac{1}{3} \le p \le 1$. To examine the entanglement of Werner state we used Peres-Horodecki criterion. The ppt-matrix for $\rho_{w,1}$ takes form
\be
\rho_{w,1}^{ppt}=\begin{pmatrix}
A & 0 & 0 & 0\\
0 & B & C & 0\\
0 & C & B & 0\\
0 & 0 & 0 & A\\
\end{pmatrix},
\ee
and has eigenvalues $\lambda_1=\frac{1-3p}{4}$, $\lambda_{2,3,4}=\frac{1+p}{4}$, which must be non-negative for separable state. Therefore $\rho_{w,1}$ describes entangled states in the domain $(1/3,1]$ of the parameter $p$.

Density matrix $\rho_{w,2}=\frac{1}{Tr[\rho_{w,1}^{2}]}\rho_{w,1}^{2}$ is obtained from matrix $\rho_{w,1}$ by the nonlinear transform. The normalized density matrix has the form similar to matrix $\rho_{w,1}$, but with redefined numbers $A(p)=\frac{(1+p)^2+4p^2}{4(1+3p^2)}$, $B(p)=\frac{(1-p)^2}{4(1+3p^2)}$, $C(p)=\frac{p(1+p)}{1+3p^2}$. Its eigenvalues are $\lambda_1=\frac{(1+3p)^2}{4(1+3p^2)}$, $\lambda_{2,3,4}=\frac{(1-p)^2}{4(1+3p^2)}$. Unlike $\rho_{w,1}$ the density matrix is non-negative for all real values of parameter p. The ppt-matrix for $\rho_{w,2}$ has eigenvalues $\lambda_1=\frac{1-6p-3p^2}{4(1+3p^2)}$, $\lambda_{2,3,4}=\frac{1+2p+5p^2}{4(1+3p^2)}$, they are non-negative if $(-1-\frac{2\sqrt{3}}{3} \le p \le -1+\frac{2\sqrt{3}}{3})$. According to Peres-Horodecki criterion $\rho_{w,2}$ describes entangled states if $(p < -1-\frac{2\sqrt{3}}{3})\cup(-1+\frac{2\sqrt{3}}{3} < p)$.

Matrix $\rho_{w,3}=\frac{1}{Tr[\rho_{w,1}^3]}\rho_{w,1}^3$ is also similar to $\rho_{w,1}$ with $A(p)=\frac{(1+p)^3+12p^2(1+p)}{4(6p^3+9p^2+1)}$, $B(p)=\frac{(1-p)^3}{4(6p^3+9p^2+1)}$, $C(p)=\frac{6p(1+p)^2+8p^3}{4(6p^3+9p^2+1)}$. For positivity of matrix $\rho_{w,3}$ it is necessary for its eigenvalues to be non-negative. First of them $\lambda_1=\frac{(1+3p)^3}{4(6p^3+9p^2+1)}$ changes sign in points $-1/3$, $-\frac{1}{2}(1+\frac{1}{\sqrt[3]{3}}+\sqrt[3]{3})$. Others three eigenvalues are equal to $\lambda_{2,3,4}=\frac{(1-p)^3}{4(6p^3+9p^2+1)}$ and they are positive if $\frac{(1-p)^3}{4(6p^3+9p^2+1)} < p < 1$. Thus $\rho_{w,3}$ is density matrix for $-1/3 \le p \le 1$. Transposed matrix has following eigenvalues: $\lambda_1=\frac{1-9p-9p^2-15p^3}{4(6p^3+9p^2+1)}$, $\lambda_{2,3,4}=\frac{1+3p+15p^2+13p^3}{4(6p^3+9p^2+1)}$, which are non-negative for $-1 \le p \le -\frac{1}{5}(1+2\frac{1}{\sqrt[3]{3}}-2\sqrt[3]{3})$. The entanglement appears for the parameter of Werner state in the domain $-\frac{1}{5}(1+2\frac{1}{\sqrt[3]{3}}-2\sqrt[3]{3})) < p \le 1$.

\section{Case of arbitrary integer $n$}
\pst
For general integer $n$ providing matrix $\rho_{w,1}^n$ we can obtain the factorized form of matrix $\rho_{w,n}$. We have the decomposition $\rho_{w,1}=SDS^{-1}$, where $D$ - diagonal matrix with eigenvalues of $\rho_{w,1}$ on the diagonal, and $S$ - unitary matrix with columns taken as eigenvectors of $\rho_{w,1}$. Using this decomposition it is easy to calculate the matrix $\rho_{w,1}^n$:
\be
\rho_{w,1}^n=(SDS^{-1})^n=SDS^{-1}\cdot SDS^{-1}\cdot\ldots\cdot SDS^{-1}\cdot SDS^{-1}=SD^nS^{-1}.
\ee
Eigenvectors of $\rho_{w,1}$ read
\be
\begin{matrix}

\lambda_1\to \overrightarrow{a_1}=(\begin{matrix}
\frac{1}{\sqrt{2}} & 0 & 0 & \frac{1}{\sqrt{2}}
\end{matrix})^T,\\
\lambda_{2,3,4}\to \begin{matrix}
\overrightarrow{a_2}=(\begin{matrix}
0 & 1 & 0 & 0
\end{matrix})^T,\\
\overrightarrow{a_3}=(\begin{matrix}
0 & 0 & 1 & 0
\end{matrix})^T,\\
\overrightarrow{a_4}=(\begin{matrix}
\frac{1}{\sqrt{2}} & 0 & 0 & -\frac{1}{\sqrt{2}}
\end{matrix})^T,\\
\end{matrix}
\end{matrix}
\ee
where $\lambda_i$ - eigenvalues (2). Thus $\rho_{w,1}$ has the following decomposition:
\be
\rho_{w,1}=SDS^{-1}=\begin{pmatrix}
\frac{1}{\sqrt{2}} & 0 & 0 & \frac{1}{\sqrt{2}}\\
0 & 1 & 0 & 0\\
0 & 0 & 1 & 0\\
\frac{1}{\sqrt{2}} & 0 & 0 & -\frac{1}{\sqrt{2}}
\end{pmatrix}
\begin{pmatrix}
\frac{1+3p}{4} & 0 & 0 & 0\\
0 & \frac{1-p}{4} & 0 & 0\\
0 & 0 & \frac{1-p}{4} & 0\\
0 & 0 & 0 & \frac{1-p}{4}
\end{pmatrix}
\begin{pmatrix}
\frac{1}{\sqrt{2}} & 0 & 0 & \frac{1}{\sqrt{2}}\\
0 & 1 & 0 & 0\\
0 & 0 & 1 & 0\\
\frac{1}{\sqrt{2}} & 0 & 0 & -\frac{1}{\sqrt{2}}
\end{pmatrix}.
\ee
Applying this decomposition, we obtain for $\rho_{w,n}$ the expression:
\be
\rho_{w,n}=\frac{1}{3(1-p)^n+(1+3p)^n}\left(\begin{smallmatrix}
\frac{1}{2}((1+3p)^n+(1-p)^n)&0&0&\frac{1}{2}((1+3p)^n-(1-p)^n)\\
0&(1-p)^n&0&0\\
0&0&(1-p)^n&0\\
\frac{1}{2}((1+3p)^n-(1-p)^n)&0&0&\frac{1}{2}((1+3p)^n+(1-p)^n)
\end{smallmatrix}\right).
\ee
Eigenvalues of this matrix read
\be
\lambda_{1}=\frac{(1+3p)^n}{3(1-p)^n+(1+3p)^n},\\
\ee
\be
\lambda_{2,3,4}=\frac{(1-p)^n}{3(1-p)^n+(1+3p)^n}.
\ee
If $n$ goes to infinity, all the eigenvalues become non-negative numbers for all the values of real parameter $p$:
\be\begin{matrix}
p <0 \to \lambda_1=0, \lambda_{2,3,4}=\frac{1}{3},\\
p =0 \to \lambda_{1,2,3,4}=\frac{1}{4},\\
p >0 \to \lambda_1=1, \lambda_{2,3,4}=0.\\
\end{matrix}
\ee
Thus, in this limit the matrix $\rho_{w,n}$ satisfies the conditions for density matrix for all real values of parameter $p$. The ppt-matrix $\rho_{w,n}^ppt$ has the following eigenvalues:
\be
\lambda_{1}^{ppt}=\frac{1}{2}\frac{3(1-p)^n-(1+3p)^n}{3(1-p)^n+(1+3p)^n},\\
\ee
\be
\lambda_{2,3,4}^{ppt}=\frac{1}{2}\frac{(1-p)^n+(1+3p)^n}{3(1-p)^n+(1+3p)^n}.
\ee

Before utilize criterion for separable state it is worth to separate cases of odd and even $n$.

For odd integers $n$, there is singularity in eigenvalues of $\rho_{w,n}$ and $\rho_{w,n}^{ppt}$ in point $p=1+\frac{4}{\sqrt[n]{3}-3}$. The eigenvalue $\lambda_{1}$ goes to zero if $p=-1/3$, $\lambda_{2,3,4}$ go to zero in point $p=1$. Therefore $\rho_{w,n}$ is non-negative for all odd $n$ for $p \in [-1/3; 1]$. The eigenvalues $\lambda_1^{ppt}$ and $\lambda_{2,3,4}^{ppt}$ goes to zero if $p=1-\frac{4}{\sqrt[n]{3}+3}$ and $p=-1$, respectively. Finally, $\rho_{w,n}^{ppt}$ is positive for $p \in [-1; 1-\frac{4}{\sqrt[n]{3}+3}]$. So the matrix $\rho_{w,n}$ describes separable state for $p \in [-1/3; 1-\frac{4}{\sqrt[n]{3}+3}]$ and entangled state in domain $(1-\frac{4}{\sqrt[n]{3}+3}; 1]$. In the limit $n\to\infty$, where integers $n$ are odd numbers, the domain of parameter $p$ providing entangled state becomes $(+0; 1]$.

For even $n$, the matrix $\rho_{w,n}$ is non-negative for all real $p$. As for $\rho_{w,n}^{ppt}$, its eigenvalues $\lambda_{2,3,4}^{ppt}$ are always positive and $\lambda_1^{ppt}$ changes sign in points $p=1-\frac{4}{\sqrt[n]{3}+3}$ and $p=1+\frac{4}{\sqrt[n]{3}-3}$. Thus, $\rho_{w,n}$ describes entangled state if $(p<1+\frac{4}{\sqrt[n]{3}-3})\cup(p>1-\frac{4}{\sqrt[n]{3}+3})$. In the limit $n\to\infty$, where integers $n$ are even numbers, the domain becomes $(p<-1)\cup(p>0)$.

Utilizing obtained eigenvalues one can calculate negativity for the case of arbitrary $n$. In the domain of parameter $p$, where $\rho_{w,n}$ describes entangled state, negativity takes value:
\be
N=\left|\frac{2(1+3p)^n}{3(1-p)^n+(1+3p)^n}\right|.
\ee
From this formula it can be easily found that negativity reaches its maximum value at the point $p=1$ for all integers $n$.
In Figure 1 the dependence of negativity on parameter $p$ is illustrated for cases of $n=1,2,3$. For the separable states negativity for density matrix $\rho_{w,n}^{ppt}$ is equal to $1$. As for the entangled states, the negativity reflects the strength on entanglement which is maximum at $p=1$.

\section{Quantum tomography for nonlinear channels transforming density matrix of Werner state}
\pst
To describe the spin states the tomographic probability distributions identified with the system states can be used~\cite{15,16}. For direction $\overrightarrow{n}\left(\theta,\psi\right)$, which is determined by the parameters of unitary matrix $u$ the tomogram for spin with density operator $\hat{\rho}$ can be obtained from formula:
\be
W(m,\overrightarrow{n})=\langle{m}|u{\hat{\rho}}u^+|{m}\rangle,
\ee
where $m$ is spin projection on the direction $\overrightarrow{n}$.
Quantum tomogram for system with density operator $\hat{\rho}$ for two spins with $j=1/2$ can be given by formula:
\be
W(m_1,\overrightarrow{a},m_2,\overrightarrow{b})=\langle{j_1,m_1,j_2,m_2}|{U\hat{\rho}U^+}|{j_1,m_1,j_2,m_2}\rangle,
\ee
where $U=u_1\otimes{u_2}$, the product being tensor product of unitary matrices
\be
u_i=\begin{pmatrix}
\cos{\left(\frac{\theta_i}{2}\right)}e^{\frac{i}{2}(\varphi_i+\psi_i)}&\sin{\left(\frac{\theta_i}{2}\right)}e^{\frac{i}{2}(\varphi_i-\psi_i)}\\
-\sin{\left(\frac{\theta_i}{2}\right)}e^{-\frac{i}{2}(\varphi_i-\psi_i)}&\cos{\left(\frac{\theta_i}{2}\right)e^{-\frac{i}{2}(\varphi_i+\psi_i)}}
\end{pmatrix}, i=(1,2).
\ee
The matrix elements of the matrices depend on Euler’s angles $\theta_i$, $\varphi_i$, $\psi_i$. For the Werner state tomographic probability for system of two qubits, depending on directions $\overrightarrow{n_1}$ and $\overrightarrow{n_2}$, read 
\begin{eqnarray}
W(\uparrow,\overrightarrow{n_1},\uparrow,\overrightarrow{n_2})=A(p)\left(\cos^2{\left(\frac{\theta_1}{2}\right)}\cos^2{\left(\frac{\theta_2}{2}\right)}+\sin^2{\left(\frac{\theta_1}{2}\right)}\sin^2{\left(\frac{\theta_2}{2}\right)}\right)+\nonumber\\+B(p)\left(\sin^2{\left(\frac{\theta_1}{2}\right)}\cos^2{\left(\frac{\theta_2}{2}\right)}+\cos^2{\left(\frac{\theta_1}{2}\right)}\sin^2{\left(\frac{\theta_2}{2}\right)}\right)+\frac{C(p)}{2}\sin\left({\theta_1}\right)\sin\left({\theta_2}\right)\cos\left(\psi_1+\psi_2\right),
\end{eqnarray}
\begin{eqnarray}
W(\uparrow,\overrightarrow{n_1},\downarrow,\overrightarrow{n_2})=A(p)\left(\sin^2{\left(\frac{\theta_1}{2}\right)}\cos^2{\left(\frac{\theta_2}{2}\right)}+\cos^2{\left(\frac{\theta_1}{2}\right)}\sin^2{\left(\frac{\theta_2}{2}\right)}\right)+\nonumber\\+B(p)\left(\cos^2{\left(\frac{\theta_1}{2}\right)}\cos^2{\left(\frac{\theta_2}{2}\right)}+\sin^2{\left(\frac{\theta_1}{2}\right)}\sin^2{\left(\frac{\theta_2}{2}\right)}\right)-\frac{C(p)}{2}\sin\left({\theta_1}\right)\sin\left({\theta_2}\right)\cos\left(\psi_1+\psi_2\right),
\end{eqnarray}
\begin{eqnarray}
W(\downarrow,\overrightarrow{n_1},\uparrow,\overrightarrow{n_2})=A(p)\left(\sin^2{\left(\frac{\theta_1}{2}\right)}\cos^2{\left(\frac{\theta_2}{2}\right)}+\cos^2{\left(\frac{\theta_1}{2}\right)}\sin^2{\left(\frac{\theta_2}{2}\right)}\right)+\nonumber\\+B(p)\left(\cos^2{\left(\frac{\theta_1}{2}\right)}\cos^2{\left(\frac{\theta_2}{2}\right)}+\sin^2{\left(\frac{\theta_1}{2}\right)}\sin^2{\left(\frac{\theta_2}{2}\right)}\right)-\frac{C(p)}{2}\sin\left({\theta_1}\right)\sin\left({\theta_2}\right)\cos\left(\psi_1+\psi_2\right),
\end{eqnarray}
\begin{eqnarray}
W(\downarrow,\overrightarrow{n_1},\downarrow,\overrightarrow{n_2})=A(p)\left(\cos^2{\left(\frac{\theta_1}{2}\right)}\cos^2{\left(\frac{\theta_2}{2}\right)}+\sin^2{\left(\frac{\theta_1}{2}\right)}\sin^2{\left(\frac{\theta_2}{2}\right)}\right)+\nonumber\\+B(p)\left(\sin^2{\left(\frac{\theta_1}{2}\right)}\cos^2{\left(\frac{\theta_2}{2}\right)}+\cos^2{\left(\frac{\theta_1}{2}\right)}\sin^2{\left(\frac{\theta_2}{2}\right)}\right)+\frac{C(p)}{2}\sin\left({\theta_1}\right)\sin\left({\theta_2}\right)\cos\left(\psi_1+\psi_2\right).
\end{eqnarray}
\begin{figure}[ht]
\bc \includegraphics[width=8.6cm]{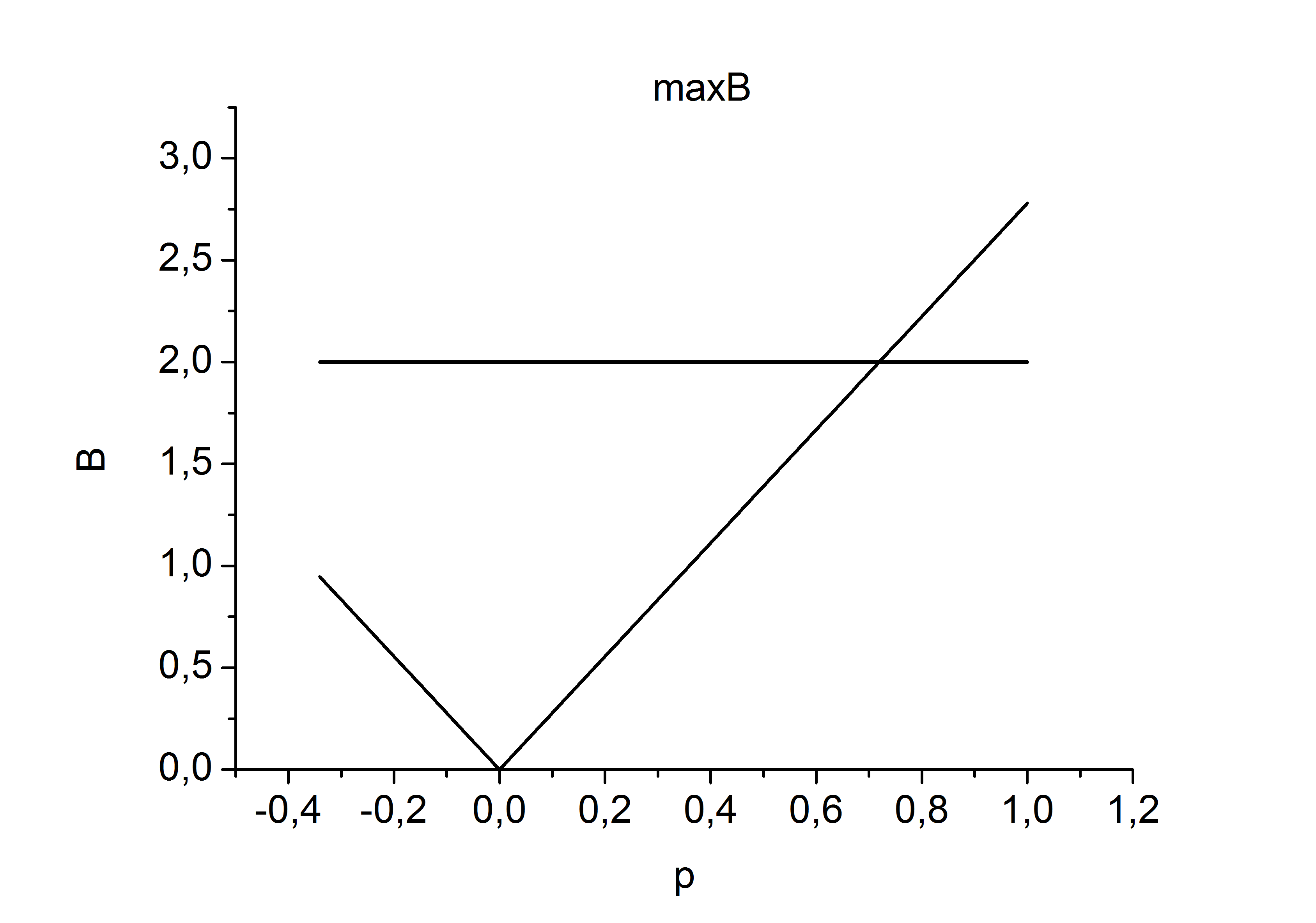}
\includegraphics[width=8.6cm]{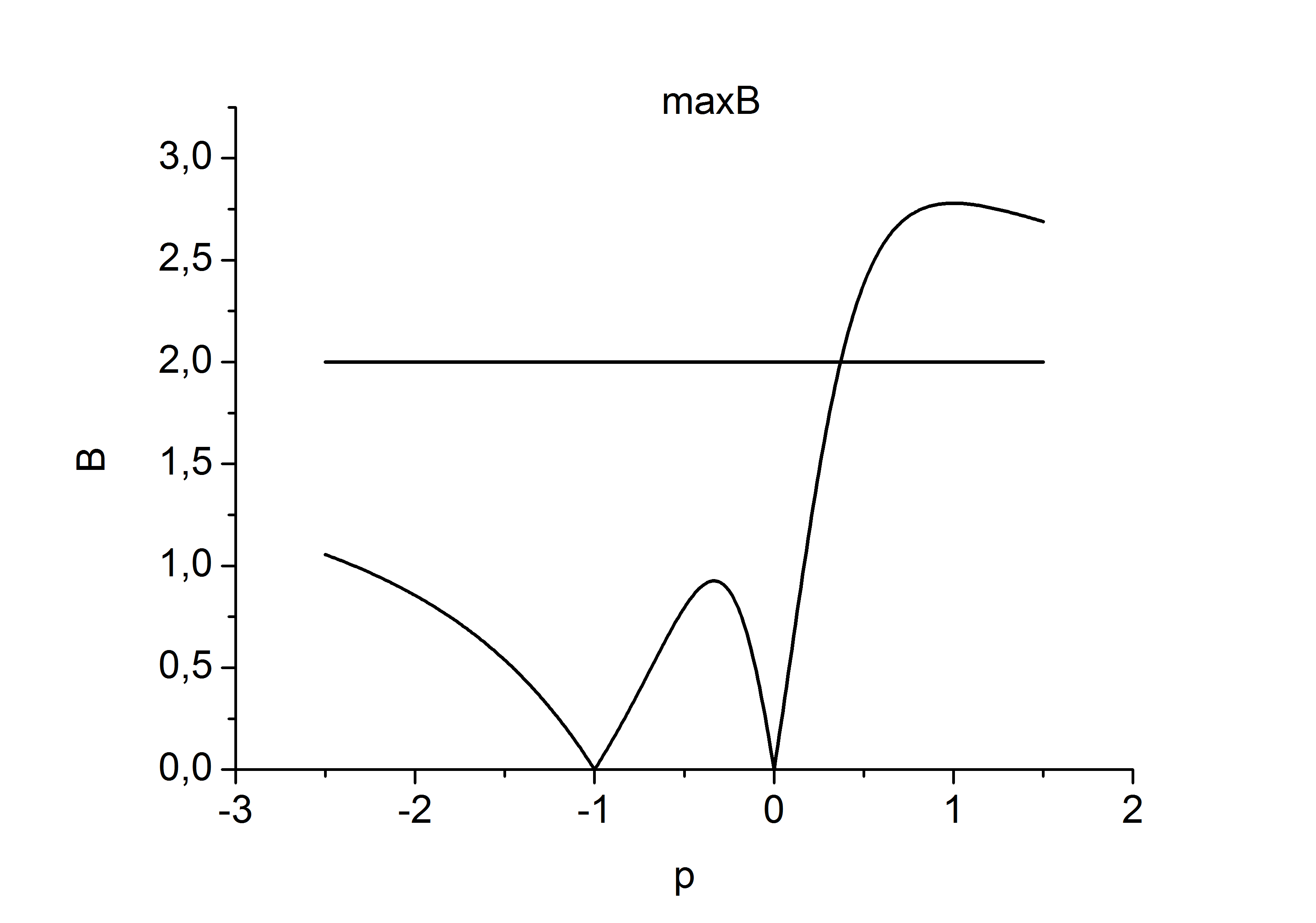}\ec
\bc
\includegraphics[width=8.6cm]{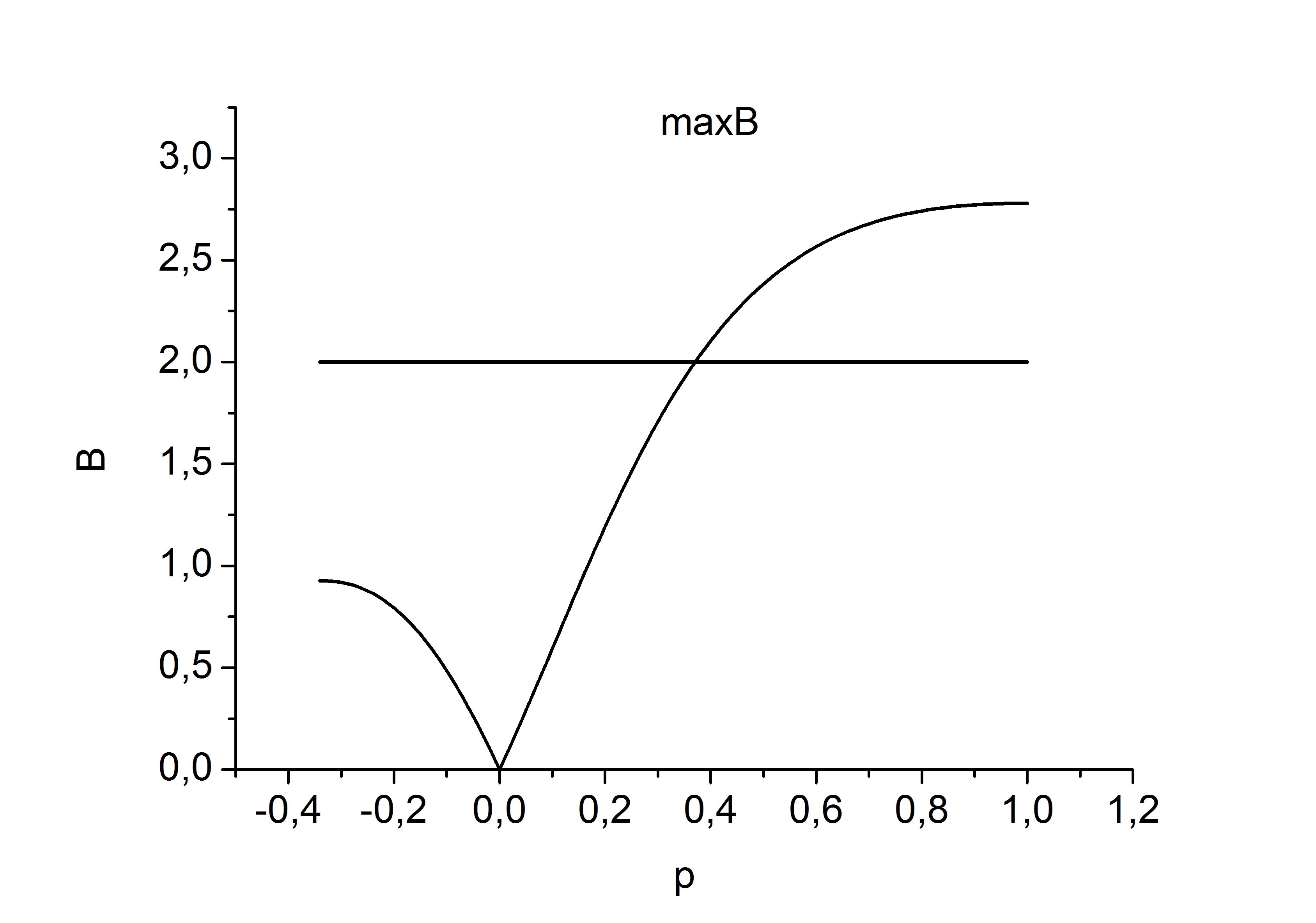}
\ec
\vspace{-4mm}
\caption{
Dependence maximum value of $\langle{M_{ab}}\rangle+\langle{M_{ac}}\rangle+\langle{M_{db}}\rangle-\langle{M_{dc}}\rangle$ on parameter $p$. Graphics correspond to cases $\rho_{w,1}$ (upper-left plot), $\rho_{w,2}$ (upper-right plot) and $\rho_{w,3}$ (lower plot). The horisontal line on all plots shows the Bell limit for separable states.} 
\end{figure}
Here $A(p)$, $B(p)$, $C(p)$ are elements of density matrix $\rho_{w,1}$. The density matrices $\rho_{w,n}$ have form similar to the form of matrix $\rho_{w,1}$. The only difference is connected with the difference of numbers $A(p)$, $B(p)$, $C(p)$, determining the matrix elements of the matrix $\rho_{w,n}$. It means that the tomogram of state with density matrix $\rho_{w,n}$ remains the same, but with following numbers:
\be
A(p)=\frac{1}{2}\frac{(1-p)^n+(1+3p)^n}{3(1-p)^n+(1+3p)^n},
\ee
\be
B(p)=\frac{(1-p)^n}{3(1-p)^n+(1+3p)^n},
\ee
\be
C(p)=\frac{1}{2}\frac{(1+3p)^n-(1-p)^n}{3(1-p)^n+(1+3p)^n}.
\ee
Having the tomogram we can examine Bell inequality~\cite{13,14,17} connected with quantum correlations in the system. For directions $\overrightarrow{n_{\alpha}}$ and $\overrightarrow{n_{\beta}}$, which we mark by indices $\alpha$ and $\beta$ the function $\langle{M_{\alpha\beta}}\rangle$ corresponding to matrices $\rho_{w,n}$ reads
\begin{eqnarray}
\langle{M_{\alpha\beta}}\rangle=W(\uparrow,\overrightarrow{n_{\alpha}},\uparrow,\overrightarrow{n_{\beta}})-W(\uparrow,\overrightarrow{n_{\alpha}},\downarrow,\overrightarrow{n_{\beta}})-W(\downarrow,\overrightarrow{n_{\alpha}},\uparrow,\overrightarrow{n_{\beta}})+W(\downarrow,\overrightarrow{n_{\alpha}},\downarrow,\overrightarrow{n_{\beta}})=\\
=2C(p)(\cos\theta_\alpha\cos\theta_\beta+\sin\theta_\alpha\sin\theta_\beta\cos\left({\psi_\alpha+\psi_\beta}\right),
\end{eqnarray}
where we used difference between elements of matrix $\rho_{w,n}$: $C(p)=A(p)-B(p)$. We obtain the expression for correlation $B$:
\begin{eqnarray}
B=\langle{M_{ab}}\rangle+\langle{M_{ac}}\rangle+\langle{M_{db}}\rangle-\langle{M_{dc}}\rangle=2C(p)(\cos\theta_a\cos\theta_b+\sin\theta_a\sin\theta_b\cos(\psi_a+\psi_b)+\nonumber\\+\cos\theta_a\cos\theta_c+\sin\theta_a\sin\theta_c\cos(\psi_a+\psi_c)+\nonumber\\+\cos\theta_d\cos\theta_b+\sin\theta_d\sin\theta_b\cos(\psi_d+\psi_b)-\nonumber\\-\cos\theta_d\cos\theta_c-\sin\theta_d\sin\theta_c\cos(\psi_d+\psi_c)).
\end{eqnarray}

As it can be seen from (25) the expression $\langle{M_{ab}}\rangle+\langle{M_{ac}}\rangle+\langle{M_{db}}\rangle-\langle{M_{dc}}\rangle$ is the product of two functions: $f(p)=2C(p)$,depending only on parameter $p$, and the function of angle arguments $\zeta(\overrightarrow{a},\overrightarrow{b},\overrightarrow{c},\overrightarrow{d})$. Thus the task of finding  maximum of $\langle{M_{ab}}\rangle+\langle{M_{ac}}\rangle+\langle{M_{db}}\rangle-\langle{M_{dc}}\rangle$ has two aspects. This function reaches its maximum when both $f(p)$ and function $\zeta(\overrightarrow{a},\overrightarrow{b},\overrightarrow{c},\overrightarrow{d})$ reach their maxima. As it can be seen from (25), the function of angles
\begin{eqnarray}
\zeta(\overrightarrow{a},\overrightarrow{b},\overrightarrow{c},\overrightarrow{d})=\cos\theta_a\cos\theta_b+\sin\theta_a\sin\theta_b\cos(\psi_a+\psi_b)+\cos\theta_a\cos\theta_c+\sin\theta_a\sin\theta_c\cos(\psi_a+\psi_c)+\nonumber\\+\cos\theta_d\cos\theta_b+\sin\theta_d\sin\theta_b\cos(\psi_d+\psi_b)-\cos\theta_d\cos\theta_c-\sin\theta_d\sin\theta_c\cos(\psi_d+\psi_c)
\end{eqnarray}
remains the same for all powers $n$ and all parameters $p$. Maximum of function $f(p)$ can be easily calculated. In the domain of values $p$ where $\rho_{w,n}$ is non-negative, $max[f(p)]=f(p)|_{p=1}=1$. So for all powers $n$ one has inequality:
\be
max(\langle{M_{ab}}\rangle+\langle{M_{ac}}\rangle+\langle{M_{db}}\rangle-\langle{M_{dc}}\rangle) \le max(\zeta(\overrightarrow{a},\overrightarrow{b},\overrightarrow{c},\overrightarrow{d})).
\ee

Figure 2 displays the dependence of correlation $B$ on parameter $p$ for different powers of power $n=1,2,3$. The correlation $B$ was calculated numerically. Taking into account the domain of $p$ where the states are separable, one can see that for these states the Bell inequality  $B \le 2$ is satisfied. Also the correlation for entangled states meets the bound $B \le 2\sqrt{2}$.

\section{Conclusion}
\pst
To resume we point out the main results of our work. We constructed and studied the nonlinear positive maps of Werner state density matrix $\rho_{w,1}\to\rho_{w,1}^n/Tr[\rho_{w,1}^n]$ for arbitrary integer $n$. It was found that for different domains of Werner state parameter $p$, the entanglement in case of both odd and even n depends on this domain. We obtained the property of violation of Bell inequality for the nonlinearly transformed states. The Cirelson bound $2\sqrt{2}$~\cite{11} was checked to be reached for transformed Werner state. The correlation between Bell inequality violation and appearing the entanglement was discussed for all integer $n$. We have shown that in the Werner state the nonlinear channel (for $n=2$) creates the entangled state from separable one in correspondence with remark of~\cite{14}. For example, the Werner state $\rho_{w,1}$ with $p=\frac{1}{6}$ is separable, but the state $\rho_{w,2}$ for this value of $p$ has negativity of entangled state.

\end{document}